\documentclass[aps, prl, twocolumn]{revtex4-2}

\usepackage{graphicx}
\usepackage{amsmath}
\usepackage{mathtools}
\usepackage{physics}
\usepackage{hyperref}

\begin{document}

\title{Oscillating chemical reactions enable communication between responsive hydrogels}

\author{Joseph J. Webber}
 \email{joe.webber@warwick.ac.uk}
\author{Thomas D. Montenegro-Johnson}%
 \email{tom.montenegro-johnson@warwick.ac.uk}
\affiliation{Mathematics Institute, University of Warwick, Coventry CV4 7AL, UK}

\date{\today}

\begin{abstract}
    Responsive hydrogels can sense environmental stimuli and respond as actuators by expelling water and changing shape. In this article, we develop theory to demonstrate that groups of responsive hydrogels can also communicate with each other, by utilising the effect of elastic deformation on chemical reaction dynamics. Specifically, we consider a system of two spatially-separated chemically responsive hydrogels suspended in a solution in which a Belousov-Zhabotinsky (BZ)-type reaction occurs. Solving for the gel dynamics with the transport of solvent through the poroelastic network and the chemical kinetics, we show how the periodic swelling-deswelling oscillations of each gel can become coupled, and how this coupling can be exploited to send signals from one gel to the other via mechanical manipulation of the sender that affect the local (and thus global) frequency of oscillation.
\end{abstract}

\maketitle

\section{Introduction}
Hydrogels are increasingly employed in the construction of biocompatible, flexible machinery \cite{lee_hydrogel_robotics_2020} that exploits their soft, elastic and porous properties. These gels comprise a cross-linked hydrophilic polymer surrounded by water molecules that are free to move, and volumetric changes of up to two orders of magnitude are observed when they are swollen in water from an initially dry state \cite{betrand_dynamics_gel_2016}. Of particular interest in the design of smart devices are responsive hydrogels, which lose their affinity for water in response to stimuli \cite{katke-diffusiophoretic_hydrogels_2024}.

Through the choice of polymer, gels can be produced that respond to temperature \cite{matsumoto_thermo_water_2018}, light \cite{samal_smart_applications_2014}, electromagnetic fields \cite{carayon_electro_field_2020}, and pH or chemical concentration \cite{blanc-collective-hydrogels-2024}. Recent experiments have illustrated how programmable shape changes can be realised in chemically-responsive hydrogels coupled to oscillating chemical reactions \cite{yoshida-self-gel-1996, levin-self_transformation_2020, nava-medina-self-shapes-2021}. These systems have been used in the lab to model processes such as the beating of cardiac tissue \cite{yoshida-self-muscle-2012}, peristaltic squeezing \cite{mao-contraction-gels-2020} and to transport droplets \cite{murase-design_gel_2008}. In addition to a wealth of experimental studies of these so-called `BZ gels' that employ the famous Belousov-Zhabotinsky reaction \cite{belousov-periodic-mechanism-1958}, there has been much progress modelling such gels \cite{kuksenok-three_reaction_2008}. Models usually treat the hydrogel as a lattice of springs whose natural length varies in the presence of chemical species that are generated and destroyed following one of a number of popular models for the BZ reaction \cite{field-oscillations_reaction_1974}.

There exist abundant examples of gels that sense the products of chemical reactions and respond accordingly, but it is also possible for these gels to influence the dynamics of the reactions themselves, enabling feedback mechanisms. \citet{nava-medina-self-shapes-2021} constructed a BZ gel with a catalyst bonded to the polymer chains such that the deformation of the gel influences the rate of reaction. These two-way relationships between a gel and its environment can be exploited to develop new paradigms for smart devices that both sense and influence their surroundings, or coordinate in systems inspired by living matter.

Chemical signalling is used in biological systems both between organisms (e.g. quorum sensing within bacterial colonies) and within multicellular life (in hormonal signals between organs) \cite{combarnous-cell_interplays_2020}. Recent studies have used BZ gels as an analogue for such processes, leading to collective motion of `swarms' of gels that communicate using diffused products \cite{teng-2022_heterogeneity_gels}. Inspired by these natural systems, we consider a pair of gels coupled to an oscillating reaction, with catalyst molecules bound to their polymer scaffolds. This system exhibits periodic oscillations in the gels, allowing signals to be sent through exploiting the diffusion of chemical solute in the solvent separating them. Signalling is achieved via mechanical manipulation of the `sender' gel, adjusting the local frequency of oscillation and hence the global dynamics in a manner akin to frequency modulation radio. Not only does this provide a model for intercellular communication, it also unlocks the development of smart hydrogel devices that coordinate complicated responses using insights from biology, employing tried-and-tested building blocks that are straightforward to implement in the real world.

\section{Modelling foundations}
We first seek a description of the response of a hydrogel to a chemical concentration field $c(\boldsymbol{x},\,t)$. This is traditionally achieved using an expression for the energy density of the deformed gel, which is then used to find stresses on virtual springs joining points in the gel -- the gel lattice-spring model of \citet{kuksenok-three_reaction_2008}.

In the present study we restrict our attention to gels undergoing uniaxial deformation and, instead of specifying a particular elastic and energetic law for the hydrogel, consider an osmotic pressure, linear in the polymer volume fraction $\phi$, that changes in nature when a critical chemical concentration $c^*$ is reached:
\begin{equation}
    \Pi = \Pi_0 \frac{\phi - \phi_0(c)}{\phi_0(c)} \qq{with} \phi_0(c) = \begin{dcases} \phi_{00} & c \le c^* \\ \phi_{0\infty} & c > c^* \end{dcases},
\end{equation}
where $\phi_0$ is the equilibrium polymer fraction \cite{etzold_transpiration_hydrogels_2021}. If $c>c^*$, the gel will deswell from an initial equilibrium $\phi \equiv \phi_{00}$, reversibly, to an equilibrium polymer fraction $\phi_{0\infty} > \phi_{00}$ in the absence of any mechanical constraints. 

Changes in $\phi$ result from the flow of water through the scaffold, which can be quantified in a uniaxial geometry by the interstitial fluid flux
\begin{equation}
    q = \frac{D(\Phi)}{\Phi}\pdv{\Phi}{x} = \frac{k \Pi_0}{\mu_l}\left(\frac{1}{\Phi_{0}} + \frac{4 \mu_s}{3\Pi_0\phi}\Phi^{\frac{1}{3}}\right)\pdv{\Phi}{x},
    \label{eqn:int_flux}
\end{equation}
where $\Phi=\phi/\phi_{00}$, $k$ is the gel permeability, $\mu_s$ is the shear modulus, $\mu_l$ is the dynamic viscosity of the water and $\Pi_0$ is a constant osmotic modulus \cite{webber-linear_hydrogels_2023a}, as derived in the Appendix. If the initial thickness of swollen hydrogel is $a_0$ and the gel occupies the space $0 \le x \le a(t)$, swelling and drying are governed by the nonlinear diffusion equation
\begin{equation}
    \pdv{\Phi}{t} = \pdv{x}\left[D(\Phi)\pdv{\Phi}{x}\right] \qq{with} \begin{dcases}\Phi(a(t),\,t) &= \Phi_0 \\ \partial_x \Phi(0,\,t) &= 0\end{dcases},
    \label{eqn:phi_evol}
\end{equation}
with the boundary condition at $x=0$ arising from no flow ($q=0$) and that at $x=a(t)$ from a stress balance \cite{webber-linear_hydrogels_2023a}, again justified in the Appendix. Polymer conservation sets the thickness of the gel in terms of its initial value,
\begin{equation}
    \int_0^{a(t)}{\Phi\,\mathrm{d}x} = a_0 \qq{so} \dv{a}{t} = -\frac{D(\Phi_0)}{\Phi_0}\partial_x\Phi(a(t),\,t).
\end{equation}

\subsection{Reaction modelling}
The best-known example of an oscillatory chemical reaction is the Belousov-Zhabotinsky (BZ) reaction with a bromate oxidiser \cite{belousov-periodic-mechanism-1958}. A number of different models for this reaction have been proposed in the literature \cite{field-oscillations_reaction_1974}, including the Brusselator \cite{glandsdorff-thermodynamic-fluctuations-1971}, featuring four reaction steps,
\begin{gather*}
    A \to X, \; 2X+Y \to 3X, \; B+X \to Y+F, \; \text{and} \;\, X \to G.
\end{gather*}
In the present study, we will consider only this model (a general approach for autocatalytic reactions) in lieu of more complex models that are specific to the BZ reaction (such as the Oregonator) owing to its analytic simplicity and the fact it retains the key mechanisms driving chemical oscillations, but this model could easily be replaced with any other approach in the governing equations. In the case of an excess of $A$ and $B$, the equations governing the reaction can be rescaled, and reaction dynamics are described by
\begin{subequations}
    \begin{align}
        \dot{X} &= r R_X(X,\,Y) = r \left[A + X^2 Y - (1+B)X\right], \\
        \dot{Y} &= r R_Y(X,\,Y) = r \left[-X^2 Y + B X\right],
    \end{align}
    \label{eqn:brusselator}%
\end{subequations}
where $X$ and $Y$ are the concentration of the two product species and $r$ is a rate constant. The assumption that $A$ and $B$ cannot be depleted allows for the existence of solutions that perpetually oscillate, with a Hopf bifurcation where $B=1+A^2$ and the single fixed point becomes unstable. These oscillations take the form of limit cycles \cite{lavenda-1971_chemical_oscillations} whose approximate period is
\begin{equation}
    T \approx (1+B)^2/4rA,
    \label{eqn:period}
\end{equation}
when $B/A \gg 1$, as calculated in the Supplementary Material through noting that the system takes the form of a relaxation oscillator. In spite of its apparent simplicity, solutions of the Brusselator system have been shown to provide good quantitative models for the processes that underpin BZ reactions \cite{lefever-1988_brusselator_same}, since many such oscillations decay very slowly in time.

\subsection{Chemomechanical oscillations of a single gel}
In experiments, \citet{yoshida-2000_phase_gels} note that the swelling--deswelling cycle (the `mechanical response') synchronises with the chemical oscillation, with zero phase difference. Though the large number of parameters in this study make fitting our results to existing data a possible yet non-unique endeavour, figure \ref{fig:full_plots} shows that this qualitative behaviour seen in a single gel is captured here, with length changes occurring in phase with chemical concentration peaks. In the Supplementary Material we discuss the nature of these oscillations, which cannot be proven to be limit cycles but exhibit near-periodicity with an approximately constant period.

\section{Two-gel systems}
\begin{figure}
    \centering
    \includegraphics{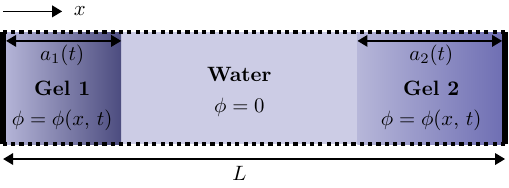}
    \vspace{-1.5em}
    \caption{A schematic illustrating the configuration of two gels in a closed container, separated by water. For clarity, the horizontal direction is shrunken, when actually we are modelling the case $a_1,\,a_2 \ll L$.}
    \label{fig:schematic}
\end{figure}
Figure \ref{fig:schematic} illustrates the model of this article, where two gels occupy the regions $0 \le x \le a_1(t)$ and $L-a_2(t)\le x \le L$, respectively, with $L \gg a_1,\,a_2$. When fully swollen in the absence of any chemical species, $a_1=a_2=a_0$ and $\Phi \equiv 1$ in both gels. Reactions occur only in the hydrogel, where there is a catalyst bonded to the polymer scaffold, and the rate of reaction is proportional to the concentration of reagents, as in the experiments of \citet{nava-medina-self-shapes-2021}. The rate constant can depend on the polymer volume fraction, since the catalyst concentration is high in drier gels where the scaffold is tightly-packed. To incorporate this effect, we replace $r$ in equation \eqref{eqn:brusselator} with $r(\Phi)$.

Gel dynamics are governed by the system \eqref{eqn:phi_evol}, where we assume $D(\Phi)\equiv D$, a constant \cite{doi-gel_dynamics_2009}. This equation can be coupled with an advection-diffusion-reaction model for the concentration of chemical species $c$,
\begin{subequations}
    \begin{align}
        \text{solute:}\quad &\pdv{c}{t}+q\pdv{c}{x} = r(\Phi) R_c(X,\,Y)+\mathcal{D}\pdv[2]{c}{x},
        \label{eqn:brusselator_spatial_variation} \\
        \text{gel:}\quad &\pdv{\Phi}{t} = D\pdv[2]{\Phi}{x}%
    \end{align}%
    \label{eqn:full_system}%
\end{subequations}
where $c$ represents $X$ or $Y$, $q=0$ outside of the gel and $\mathcal{D}$ is the molecular diffusion coefficient, equal to $D_w$ in the water and $D$ in the gel. We assume that the gels respond to the concentration of species $Y$, with rapid deswelling occurring above a critical threshold $Y = Y^*$.

\begin{figure}
    \centering
    \includegraphics{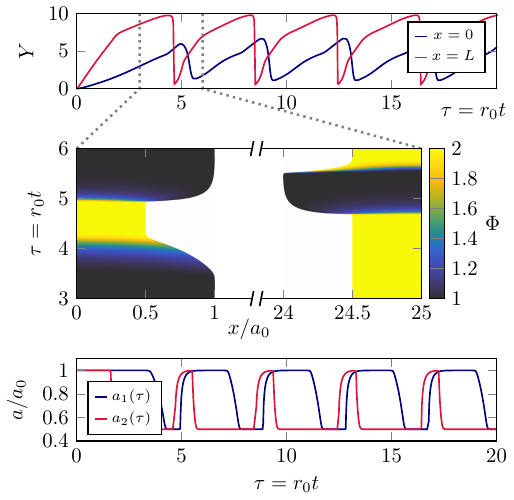}
    \caption{Plots of the length of the two gels, profiles of the gel composition and the concentration of the product $Y$ at $x=0$ and $x=L$ showing how the two gels shrink and reswell. In this plot, $A=1$, $B=5$ and $L=25a_0$. The rate constant $r(\Phi)=r_0$ in the left-hand gel and $5r_0$ in the right-hand gel. The Damk\"{o}hler number $Da_I = 1$ with $t_{\text{diff}} = 0.125 t_{\text{pore}}$. The critical concentration $Y^* = 5$ and $\phi_{0\infty} = 2\phi_{00}$.}
    \label{fig:full_plots}
\end{figure}

Figure \ref{fig:full_plots} shows an example solution to this coupled set of partial differential equations, illustrating how the shape of a gel and its composition both change in an oscillating pattern in response to chemical signals, which diffuse across the gap between gels. It is difficult to determine whether the oscillations seen here are limit cycles or evidence of quasiperiodicity, but it is clear from figure \ref{fig:full_plots} and justified in the Supplementary Material that at least near-periodic solutions persist for long times.

\subsection{Coupled oscillator model}
It is difficult to make predictions from this system of equations, and some effects, such as the advective term on the left-hand side of equations \eqref{eqn:brusselator_spatial_variation}, are less important than others: e.g. the reaction dynamics or the diffusion of species between hydrogels. There are three timescales in the problem, corresponding to chemical kinetics, diffusion in the water, and the poroelastic transport in the gels,
\begin{equation}
    t_{\text{react}} = 1/r_0,\; t_{\text{pore}} = \mu_l a_0^2/k \Pi_0, \, \text{and} \, t_{\text{diff}} = L^2/D_w,
\end{equation}
where $r_0$ is a typical scale for $r(\Phi)$. In many cases, it is possible to reduce the generally-applicable full system \eqref{eqn:full_system} to a system of coupled oscillators. This behaviour is well-attested experimentally in studies of coupled BZ droplets, where rich coupled-oscillator dynamics have been observed \cite{li-2014_combined_droplets}.

Such a system of `diffusively-coupled' Brusselators has been studied in the literature \cite{volkov-1995_bifurcations_brusselators,drubi-2007_coupling_chaos,mayora-cebollero-2025_almost_systems}, where two oscillators are linked by the transport of chemical species between them. The analysis of these systems is mathematically complicated, with the existence, stability and dynamics of oscillatory solutions dependent on the strength of coupling, as discussed in the Supplementary Material. In order to deduce the coupling strength in our system, we must parameterise the diffusive transport of chemical species across the gap separating the two BZ gels. 

Firstly, we assume that $t_{\text{react}} \gg t_{\text{pore}}$. This latter timescale is set by diffusion through pores, and, though this is typically of the order of minutes, the reaction timescale can be tuned to be slower by setting the amount of catalyst bonded to the gel. Scaling time with $t_{\text{react}}$ in equations \eqref{eqn:brusselator_spatial_variation} shows that the chemical concentration is approximately uniform within the gels, as illustrated in equation \eqref{eqn:app:scaling_gel} in the appendix. We now denote the concentrations in the two gels by $X_1$, $X_2$, $Y_1$ and $Y_2$.

The ratio $t_{\text{pore}}/t_{\text{react}}$ is equal to the first Damk\"{o}hler number $Da_I$, and taking $Da_I \ll 1$ suggests that the gel reconfigures rapidly to changes in chemical concentration, instantaneously reaching $\Phi_0(Y)$, allowing us to neglect the influence of gel dynamics on the problem. In the water, scaling times with $t_{\text{react}}$, we assume that diffusion occurs on a much faster timescale than the reaction or diffusion in the gel, $t_{\text{diff}} \gg t_{\text{pore}} \gg t_{\text{react}}$, and so, as seen in equation \eqref{eqn:app:scaling_water}, the concentration fields are linear between the two hydrogels. 

Thus, the flux of chemical species across the gel--water boundary can be approximated by the diffusive flux $Q_c$ of chemical species $c$ within the water,
\begin{equation}
    Q_c = -D_w \pdv{c}{x} = -\frac{D_w (c_2-c_1)}{L-a_1-a_2} \approx -\frac{D_w (c_2-c_1)}{L},
\end{equation}
assuming that $a_0 \ll L$. Homogeneity within the gels implies that the chemical species delivered to the gel spreads out instantaneously, increasing the concentration at a rate $-Q_c/a_1$ in the left-hand gel and $Q_c/a_2$ in the right-hand gel. Finally, we introduce a non-dimensional time $\tau = t/t_{\text{react}}$, and assume the simplest possible form for $r(\Phi) = r_0\Phi$, to give a system of coupled equations for the concentration $c$ of $X$ or $Y$ in each gel,
\begin{equation}
    \dv{c_i}{\tau} = \Phi_i R_c + \frac{Q a_0}{a_i}(c_j-c_i), \quad Q = \frac{L}{a_0}\frac{t_{\text{react}}}{t_{\text{diff}}} \gg 1,
    \label{eqn:brusselator_coupled}
\end{equation}
where $i=1,\,2$, $i \ne j$ and $\Phi_i = \phi_0(Y_i)/\phi_{00}$.

\subsection{Linking gel mechanics to reaction dynamics}
The dynamics of the system of equations \eqref{eqn:brusselator_coupled} can be influenced by varying the reaction rate in one gel but not the other. We achieve this by imposing a mechanical strain on the second gel, such that we can control the concentration of catalyst in the reaction occurring there. Applying a uniform vertical compressive strain $E$ to the gel changes the horizontal stress balance at the interface with water, and hence changes the polymer fraction in both swollen and deswollen states \footnote{A compressive vertical strain $E$ gives a deviatoric vertical strain $\epsilon_{zz} = \Phi^{1/3}-1-E$, and the deviatoric strains in the directions parallel to the walls are equal $\epsilon_{xx}=\epsilon_{yy}$. Thus, $\epsilon_{xx}=-\epsilon_{zz}/2$.},
\begin{equation}
    \frac{\Phi-\Phi_0(c)}{\Phi_0(c)} = \frac{\mu_s}{\Pi_0}\left(1+E-\Phi^{1/3}\right).
    \label{eqn:phi0_compress}
\end{equation}
Therefore, assuming there are no mechanical constraints on the left-hand gel as it shrinks and swells,
\begin{equation}
    \Phi_1 = \begin{dcases}1 & Y_1 \le Y^* \\ \Phi_{0\infty} & Y_1 > Y^*\end{dcases} ;\; \Phi_2 = \begin{dcases}\Phi_2^{-} & Y_2 \le Y^* \\ \Phi_2^+ & Y_2 > Y^*\end{dcases},
\end{equation}
with $\Phi_2^{\pm}$ obtained by solving equation \eqref{eqn:phi0_compress} below and above $Y_2=Y^*$. The gel widths are determined by conservation of polymer,
\begin{equation}
    a_1 = \Phi_1^{-1}a_0 \qq{and} a_2 = (1-E)^{-1}\Phi_2^{-1}a_0.
    \label{eqn:ai}
\end{equation}

\begin{figure}
    \centering
    \includegraphics{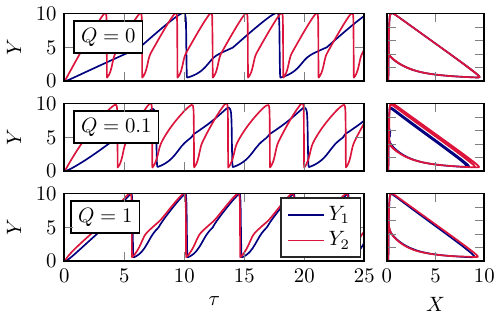}
    \caption{Solutions to the simplified coupled oscillator model \eqref{eqn:brusselator_coupled} for different coupling strengths $Q$
    when $A=1$, $B=5$. Here, $\phi_{0\infty}=2\phi_{00}$, $Y^*=5$ and the second gel is compressed with $E=1$ and $\mu_s/\Pi_0=100$. On the right, phase-plane plots of $Y$ against $X$ showing behaviour close to periodicity when $X_i=Y_i=0$.}
    \label{fig:vary_q}
\end{figure}
Solving equations \eqref{eqn:brusselator_coupled} shows the existence of periodic behaviour, just as in the single-oscillator case. Figure \ref{fig:vary_q} illustrates how changing the coupling strength leads to synchronisation of the two nonlinear oscillators. When $Q=0$, both reactions occur independently, with the two systems exhibiting different periods. As the strength of coupling increases, the periods synchronise as integer multiples of each other before equalising at an intermediate value between the natural periods of each oscillator. The process governing this coupling for intermediate $Q$ is explained in the Supplementary Material.

In the large-$Q$ regime, it is apparent that $X_1=X_2$ and $Y_1=Y_2$, and so equations \eqref{eqn:brusselator_coupled} reduce to those for a single oscillator of the form of equation \eqref{eqn:brusselator} with rate $r=(\Phi_1+\Phi_2)/2$. We can therefore use an analogous approach to that for the single oscillator \eqref{eqn:period} to find the oscillation period,
\begin{align}
    T\!=\! \frac{(1+B)^2}{2 A^2 (\Phi_\infty+\Phi_2^+)}\!+\!\frac{2 Y^*}{A}\frac{(\Phi_\infty-1) + (\Phi_2^+-\Phi_2^-)}{(1+\Phi_2^-)(\Phi_\infty+\Phi_2^+)}\!\!\!
    \label{eqn:coupled_period}
\end{align}
provided that $(1+B)^2/4A > Y^*$, as also derived in the Supplementary Material.

Strong coupling occurs when the diffusivity in the water is high, reaction rate constant sufficiently low, and when the separation between gels is not too large. In general, $D_w \sim 10^{-9}\,\mathrm{m}^2\mathrm{s}^{-1}$ \cite{vignes-diffusion_composition_1966}, and $r_0$ is approximated well by the reciprocal of the characteristic timescale for the reaction. Therefore, $Q \gg 1$ when
\begin{equation}
    \text{reaction timescale} \gg \left(10^9\,\mathrm{m^{-2}}a_0 L\right) \, \mathrm{s}.
\end{equation}
For millimetre-scale systems, this requires reactions to occur on a timescale of minutes, or hours in the case of centimetre-scale systems, both reasonable choices seen in experiments \cite{levin-self_transformation_2020}.

This allows for a signal to be transmitted between two spatially-separated gels by modulating the frequency of the first gel's oscillations through varying the compression $E$ in time. In the limit of strong coupling, equation \eqref{eqn:coupled_period} gives the oscillation period that both gels settle to, and this is seen to be dependent on the polymer fraction $\Phi_2^{\pm}$, allowing measurements of the behaviour of the left-hand gel to be used to deduce the state of the right-hand gel.

\begin{figure}
    \centering
    \includegraphics{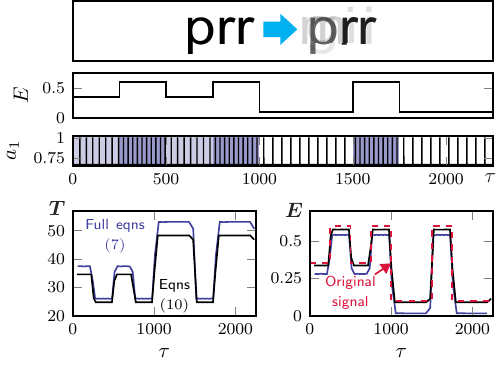}
    \caption{To send the text \texttt{PRR}, the letters are converted to values $\left\lbrace 16,\,18,\,18\right\rbrace$ which are then given ternary representations $\left\lbrace 121,\,200,\,200\right\rbrace$. Each ternary digit sets $E(\tau)$, defined piecewise in sections of length $\Delta \tau = 250$, and we can solve the simplified oscillator model \eqref{eqn:brusselator_coupled} and the full governing equations \eqref{eqn:full_system} to find the effect on the period of gel oscillation. To produce this plot, $A=1$, $B=15$, $Y^*=5$ and $\Phi_\infty=1.5$. When solving the full equations, we take $Da_I = 1$, $L=25a_0$ with $t_{\text{diff}} = 0.125 t_{\text{pore}}$. This leads to $Q=40$, used in the solution of equation \eqref{eqn:brusselator_coupled}.}
    \label{fig:send_signal}
    \vspace{-1.5em}
\end{figure}

For the sake of analytic simplicity, we restrict our attention to stiff gels with $\mu_s/\Pi_0 \gg 1$, where equation \eqref{eqn:phi0_compress} implies that $\Phi_2^\pm$ are both $(1+E)^3$. Thus, we can use equation \eqref{eqn:coupled_period} to deduce the compressive strain $E$ on the second gel from a measurement of the period $T$. If the imposed strain changes sufficiently slowly that the oscillations in both gels settle to a steady period before $E$ changes again, this can be used to send information from one gel to another through a frequency-modulated signal, as illustrated in figure \ref{fig:send_signal}. 

Here, the strain $E(\tau)$ on the second gel is varied and the time period of oscillations in the first gel's size is measured. This signal is reproduced near-perfectly from a solution to the gel system modelled by coupled oscillators, as would be expected given that our expression for oscillation period is based on this model, but perhaps more importantly this same expression, applied to a numerical solution of the full system of governing equations \eqref{eqn:full_system}, is seen to give an accurate prediction of the forcing signal. 

Small discrepancies arise due to the fact that the strongly-coupled oscillators model neglects small differences in concentration between the two gels, explaining why the black curve in figure \ref{fig:send_signal} does not exactly match the signal. Larger differences occur between the signal predicted from the solution to the full equations, resulting from gel reconfiguration, spatial variation in concentration within the hydrogels, and a nonlinear concentration profile, all of which we have neglected here. In spite of this, it is clear that the oscillator model allows for the strain $E(\tau)$ to be deduced accurately through a measurement of the response period, and can be applied to deduce this information without having to solve for the full dynamics.

\section{Conclusions}
In this article, we have introduced and modelled a frequency modulation paradigm for communication between responsive hydrogels, creating a basis for the development of increasingly complex gel machines with separate components. By exploiting the influence of mechanical manipulation on the dynamics of an oscillating chemical reaction occurring in a gel we showed how local changes in one hydrogel can affect the global dynamics of a system of oscillating gels that are coupled through diffusion of chemical species.

Much as organ systems can send out chemical signals in the form of hormones in multicellular life, or colonies of microorganisms can inform others of danger or resources \cite{combarnous-cell_interplays_2020}, this chemical signalling allows us to build up complex hydrogel machines that comprise multiple communicating parts. Mechanical input can be converted to chemical signals that then result in morphological changes, allowing for physical strains to be transmitted across a distance, in much the same way as a traditional telephone converts sound waves into an electrical signal and back again.

This has two-fold importance: firstly, it illustrates how artificial systems constructed from hydrogels can act as analogues for important biological processes such as quorum sensing, and collective dynamics can result from an external input. For example, if one gel is placed under strain, others nearby can receive this information and move away or respond accordingly. Secondly, we can use these communication processes to chain together hydrogels to achieve complex goals -- one immediate example could be coordinating a number of responsive hydrogel pumps placed at different positions to maintain a constant concentration of chemical species.

\begin{acknowledgments}
    \textit{This work was supported by the Leverhulme Trust Research Leadership Award `Shape-Transforming Active Microfluidics’ (RL-2019-014) to TDMJ. We are thankful to three anonymous referees for their helpful comments that have improved this work.}
\end{acknowledgments}

\appendix
\section{Derivation of the gel model}
\label{app:lens}
In this study, we use a linear-elastic-nonlinear-swelling (LENS) model to describe the hydrogel dynamics, basing our description of chemically-responsive gels on a recent description of temperature responsive gels \cite{webber-tubular-LENS-2024}. However, all models of poroelasticity are built on a foundation of conservation of the two components of the system -- the solid phase (in this case the polymer chains making up the gel scaffold) and the liquid phase (the water in the pore spaces) \cite{betrand_dynamics_gel_2016}. In this case, we assume that both of these phases are individually incompressible and that $\phi$ represents the local polymer volume fraction and $\phi_w = 1-\phi$ is the local water volume fraction, so that
\begin{equation}
    \pdv{\phi}{t} + \pdv{x} \left(\phi u_p\right) = 0 \; \text{and} \; \pdv{\phi_w}{t} + \pdv{x} \left(\phi_w u_w\right) = 0,
    \label{eqn:app:conservation}
\end{equation}
where $u_p$ is the polymer velocity and $u_w$ is the fluid velocity. Adding these two equations shows that
\begin{equation}
    \phi u_p + \phi_w u_w = \text{constant}.
    \label{eqn:app:divq}
\end{equation}
If, as in the geometry considered in figure \ref{fig:schematic}, there is a rigid boundary, neither water nor polymer can pass and so the average velocity of the two phases is zero: if polymer moves in one direction, water must move to fill the space it has left, and vice versa. Equation \eqref{eqn:app:divq} shows that the same assumption must be applied throughout the gel, and so $\phi u_p = -\phi_w u_w$. Defining $q = \phi_w(u_w-u_p)$ as the relative flux of water molecules against a reconfiguring scaffold, equation \eqref{eqn:app:conservation} also shows that
\begin{equation}
    \pdv{\phi}{t} = \pdv{x}\left(\phi q\right).
\end{equation}

The deswelling and reswelling behaviour of a chemically-responsive hydrogel when the external chemical field is changed occur as the polymer chains reconfigure in the presence of the solute species and their affinity for water molecules is accordingly modified. In the case of BZ gels, it is often the concentration of $[\mathrm{Ru}(bpy)_3]^+$ ions that modifies the structure of the polymer chains and leads to changes in the equilibrium degree of hydration \cite{kuksenok-three_reaction_2008}. The simplest way to encode this is through an equilibrium polymer fraction $\phi_0$ that depends on the local concentration $c$, i.e.
\begin{equation}
    \phi_0(c) = \begin{cases}\phi_{00} & c \le c^* \\ \phi_{0\infty} & c > c^*\end{cases}.
\end{equation}
A gel's propensity to swell or deswell is governed by the osmotic pressure $\Pi$, and in the absence of any external mechanical constraints, we expect $\phi \to \phi_0$ after water has been expelled or imbibed, and hence $\Pi = 0$ at the equilibrium polymer fraction. In the present study, we therefore choose a linear osmotic pressure $\Pi = \Pi_0(\phi-\phi_0)/\phi_0$ that can be expected to fit the behaviour close to the equilibrium polymer fraction well \cite{etzold_transpiration_hydrogels_2021}, and give a good qualitative description of swelling and deswelling behaviour.

The transient evolution of polymer fraction when $c$ is changed is governed by the interstitial fluid flow $q$. It is standard practice in poroelasticity to link $q$ to gradients in pore or `pervadic' \cite{peppin-pressure_suspensions_2005} pressure $p$ through Darcy's law, such that
\begin{equation}
    q = -\frac{k}{\mu_l}\pdv{p}{x},
    \label{eqn:app:darcy}
\end{equation}
where $k$ is the permeability and $\mu_l$ the dynamic viscosity of water. These gradients in pore pressure must balance gradients in bulk (effective) stress in the gel material itself such that the entire system is in force-free equilibrium, i.e.
\begin{equation}
    \pdv{x}(\sigma_{\text{eff}}-p) = 0.
\end{equation}
There are a number of different models for the effective stress in a hydrogel: in the absence of any elastic stresses, we would expect $\sigma_{\text{eff}}$ to only have a contribution from osmotic pressure $\Pi$, but adding in the effects of polymer chain elasticity leads to models ranging from the fully linear approach of \citet{doi-gel_dynamics_2009} to hyperelastic models \cite{betrand_dynamics_gel_2016,cai-equations_gels_2012}. In the present study, we have used the linear-elastic-nonlinear-swelling (LENS) theory that permits large deviations from equilibrium polymer fraction but linearises around small deviatoric strains, as introduced by \citet{webber-linear_hydrogels_2023a} and  \citet{webber-linear_hydrogels_2023b} for its analytic simplicity and good agreement with fully-nonlinear approaches using only a small number of macroscopically-meaningful parameters \cite{webber-tubular-LENS-2024}. In this case,
\begin{equation}
    \pdv{\sigma_{\text{eff}}}{x} = \frac{1}{\phi}\pdv{\Pi}{x} + \frac{4\mu_s}{3\phi}\left(\frac{\phi}{\phi_{00}}\right)^{1/3}\pdv{\phi}{x},
\end{equation}
where $\mu_s$ is the shear modulus, assumed independent of $\phi$ (as is indeed the case in many nonlinear models \cite{cai-equations_gels_2012,webber-tubular-LENS-2024}). This expression, combined with \eqref{eqn:app:darcy}, leads to the interstitial fluid flux expression of equation \eqref{eqn:int_flux}.

At the interface between gel and water, we require a boundary condition arising from stress balances. The pervadic pressure is continuous across phase boundaries \cite{peppin-pressure_suspensions_2005} and is the only component of the stress tensor in the fluid domain away from the gel. Therefore, imposing continuity of normal stress requires $\sigma_{\text{eff}} = 0$ at $x=a(t)$. In the absence of any deviatoric strains, the effective stress just has a contribution from the osmotic pressure, forcing $\Pi=0$ at the interface and setting $\phi=\phi_0$ here, whilst adding deviatoric strains (for example, by compressing the hydrogel) changes the balance and gives a drier interfacial polymer fraction, as seen in equation \eqref{eqn:phi0_compress}.

\section{Non-dimensionalisation and reduction of the full governing equations}
\begin{figure}
    \centering
    \includegraphics{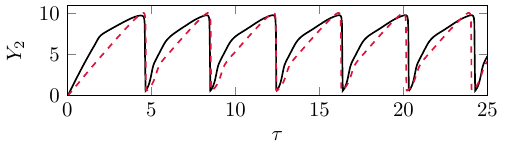}
    \caption{Plots of $Y$ evaluated at $x=L$ (solid black line) and $Y_2$ (dashed red line) in the full model of equation \eqref{eqn:full_system} and the coupled ODE system of equation \eqref{eqn:brusselator_coupled}, respectively. The same parameters are used as in figure \ref{fig:full_plots}, and we therefore take $Q = 40$ to produce the approximate plot.}
    \label{fig:compare_full_model}
\end{figure}
Equations \eqref{eqn:full_system} can be non-dimensionalised by introducing a time variable $\tau = r_0 t$ scaled with the characteristic reaction timescale $t_{\text{react}}=1/r_0$, and, scaling all lengths with $a_0$, we let $\tilde{x}=x/a_0$. Thus, recalling that $q=(D/\Phi)\pdv*{\Phi}{x}$, equation \eqref{eqn:brusselator_spatial_variation} becomes
\begin{equation}
    \pdv{c}{\tau} + \frac{D}{r_0 a_0^2\Phi}\pdv{\Phi}{\tilde{x}}\pdv{c}{\tilde{x}} = \frac{r(\Phi)}{r_0}R_c(X,\,Y) + \frac{\mathcal{D}}{r_0 a_0^2}\pdv[2]{C}{\tilde{x}}.
\end{equation}
Since $t_{\text{pore}} = a_0^2/D$ and $t_{\text{react}} = 1/r_0$, in the gel
\begin{equation}
    \pdv{c}{\tau} + \frac{t_{\text{react}}}{t_{\text{pore}}\Phi}\pdv{\Phi}{\tilde{x}}\pdv{c}{\tilde{x}} = \frac{r(\Phi)}{r_0}R_c(X,\,Y) + \frac{t_{\text{react}}}{t_{\text{pore}}} \pdv[2]{c}{\tilde{x}}.
    \label{eqn:app:scaling_gel}
\end{equation}
When $t_{\text{react}} \gg t_{\text{pore}}$, the only admissible balances in this equation involve balancing advection and diffusion, and it is clear that we expect reaction dynamics to play a role, so we must have $\pdv*{c}{\tilde{x}} = 0$ at leading order. In the water,
\begin{equation}
    \pdv{c}{\tau} = \frac{L^2}{a_0^2}\frac{t_{\text{react}}}{t_{\text{diff}}} \pdv[2]{c}{\tilde{x}},
    \label{eqn:app:scaling_water}
\end{equation}
and hence $\pdv*[2]{c}{\tilde{x}} = 0$ since $t_{\text{react}} \gg t_{\text{diff}}$ and $L \gg a_0$. Gel dynamics follow with
\begin{equation}
    \pdv{\Phi}{\tau} = \frac{t_{\text{react}}}{t_{\text{pore}}}\pdv[2]{\Phi}{\tilde{x}}.
\end{equation}
The assumption $t_{\text{react}} \gg t_{\text{pore}}$ shows that $\Phi \equiv \Phi_0(Y)$ in both gels. The coupled system of oscillators reached in equation \eqref{eqn:brusselator_coupled} can be compared with the full numerical solutions to equations \eqref{eqn:full_system}, as illustrated in figure \ref{fig:compare_full_model}, showing strong agreement with the full numerical result, capturing the period of oscillations accurately but the transient behaviour differs significantly owing to reconfiguration of the gel, not captured in the reduced model. 

%

\end{document}


\maketitle

The Brusselator dynamical system has been studied in great detail since its introduction, with many authors focusing on the relaxation oscillations that occur in particular parameter regimes \citep{lefever-1971_chemical_oscillations,lavenda-1971_chemical_oscillations, grasman-1987_asymptotic_applications}. Here, we rederive the calculation of the dynamics of oscillating solutions to the Brusselator and apply the insight from the single-reactor case to diffusively-coupled oscillators to understand the behaviour of the four-dimensional dynamical system as $Q$ is varied.

\section{Hopf bifurcation and relaxation oscillations in \texorpdfstring{$\mathbb{R}^2$}{R2}}
The Brusselator is described by the two-dimensional system of equations
\begin{equation}
    \dv{X}{t} = A + X^2 Y - (1+B) X \qq{and} \dv{Y}{t} = B X - X^2 Y,
    \label{eqn:2d_brusselator}
\end{equation}
which admits a fixed point at $(X,\,Y) = (A,\,B/A)$. The Jacobian matrix at this fixed point is
\begin{equation}
    J = \begin{pmatrix} B-1 & A^2 \\ -B & -A^2\end{pmatrix} \qq{with} \det J = A^2 \qq{and} \tr J = B - (1+A^2),
\end{equation}
so the fixed point is only unstable when $B > 1 + A^2$. It is well-known that this change in stability corresponds to a Hopf bifurcation wherein two complex conjugate eigenvalues cross the imaginary axis and a periodic solution arises \citep{grasman-1987_asymptotic_applications}. In the present study, we will only consider cases where $B$ is sufficiently large for the fixed point to be unstable and for the nature of this periodic solution to be that of a \textit{relaxation oscillator}. If $B \gg 1$, we rewrite the governing equations \eqref{eqn:2d_brusselator} as
\begin{equation}
    \varepsilon\dv{X}{t} = \varepsilon A + \varepsilon X^2 Y - (1+\varepsilon) X \qq{and} \varepsilon\dv{Y}{t} = X - \varepsilon X^2 Y,
    \label{eqn:2d_brusselator_rescaled}
\end{equation}
where $\varepsilon \ll 1$. Dynamics are therefore especially fast unless we are travelling along the $X$ or $Y$ nullclines, defined by
\begin{subequations}
    \begin{align}
        Y = \frac{1+1/\varepsilon}{X} - \frac{A}{X^2} \qq{where} &\dv{Y}{t} = A - X \quad \text{or} \\
        Y = \frac{1}{\varepsilon X} \qq{where} &\dv{X}{t} = A - X.
    \end{align}
\end{subequations}
Figure \ref{fig:phase_plot} shows an example phase plot, illustrating how trajectories collapse onto these two nullclines and elsewhere $\dv*{X}{t} \approx - \dv*{Y}{t}$, with fast dynamics. This is discussed in more detail by, for example, \citet{lavenda-1971_chemical_oscillations}. Starting at the origin $(X,\,Y)=(0,\,0)$ for illustration, a relaxation oscillation therefore follows the pattern:
\begin{figure}
    \centering
    \input{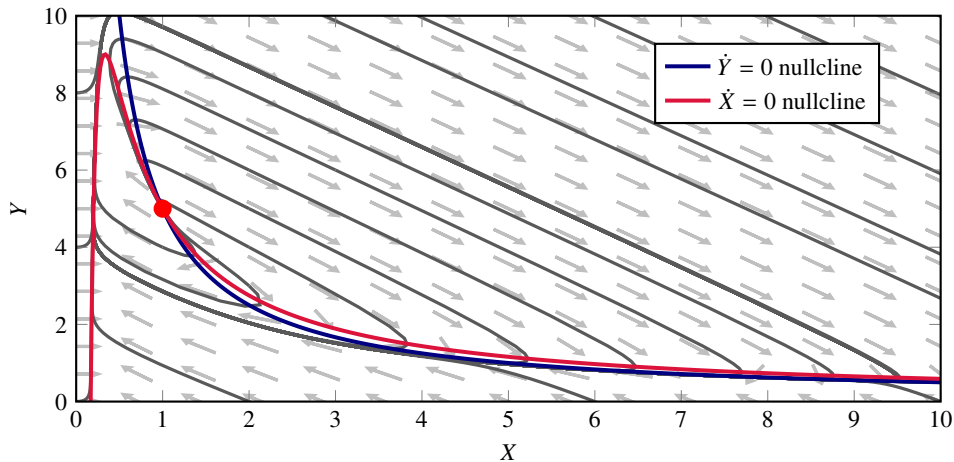}
    \caption{A phase plot of the simple uncoupled Brusselator model with $A=1$, $B=5$, showing the fixed point (in red), both nullclines and some sample trajectories in black. The slowest dynamics occur along the nullclines with rapid jumps between branches.}
    \label{fig:phase_plot}
\end{figure}

\begin{enumerate}
    \item $\dv*{X}{t}>0$ and so we immediately travel in the positive $X$ direction until the $X$ nullcline is reached. This nullcline is stable and so we travel near-vertically along it with $\dot{Y} = A - X$. The speed on this nullcline is $\mathcal{O}(1)$, comparatively slow relative to other regions.
    \item Eventually, the turning point on the nullcline is reached, at $X \approx 2\varepsilon A$, $Y \approx (1+\varepsilon)^2/(4 \varepsilon^2 A)$, where the trajectory must leave this path (to remain on it, we would require a negative $Y$ derivative) and travel rapidly along a route where $\dv*{X}{t} \approx - \dv*{Y}{t}$.
    \item This path intersects the $Y$ nullcline at $X \approx (1+\varepsilon)^2/(4 \varepsilon^2 A)$ and travels along this nullcline in the negative $X$ direction. Eventually this path is no longer stable, at the point $X = Y = \varepsilon^{-1/2}$, and the trajectory rapidly rejoins the $X$ nullcline.
    \item Since the dynamics joining the $Y$ nullcline path to the $X$ nullcline are again rapid and follow $\dv*{X}{t} \approx - \dv*{Y}{t}$, we rejoin this section of the trajectory at approximately $Y = 2 \varepsilon^{-1/2}$.
\end{enumerate}

It is immediately clear from the phase plot that this periodic cycle is unique and that the trajectories alongside the nullclines are stable, so this behaviour describes limit cycle dynamics. This was shown formally by \citet{lefever-1971_chemical_oscillations} by rewriting the system as a single second-order differential equation in the variable $\xi$, where $\xi = 1/X-1/A$, finding that
\begin{equation}
    \dv[2]{\xi}{t} + \left[\frac{A^2}{(A\xi+1)^2} + 2A\xi - (B-1)\right]\dv{\xi}{t} + \frac{A^2\xi}{1+A\xi} = 0,
\end{equation}
a specific form of Li\'enard equation for which it can be proven there must exist a limit cycle solution. 

Furthermore, the separation of timescales allows us to compute an expression for the period of such a cycle, with leading-order contributions arising only from the trajectories along nullclines where dynamics are slower. Thus,
\begin{align}
    T &\approx \int_{2/\varepsilon^{1/2}}^{(1+\varepsilon)^2/4\varepsilon^2 A}{\left(\dv{Y}{t}\right)^{-1}\,\mathrm{d}Y} + \int_{(1+\varepsilon)^2/4\varepsilon^2 A}^{1/\varepsilon^{1/2}}{\left(\dv{X}{t}\right)^{-1}\,\mathrm{d}X} \notag \\
    & \approx \frac{(1+\varepsilon)^2}{4\varepsilon^2 A^2},
    \label{eqn:time_period_calc}
\end{align}
with the logarithmic correction arising from the second integral small and vanishing as $\varepsilon \to 0$. Thus, the limit cycle can be described simply by
\begin{equation}
    Y = At \qq{and} X = \varepsilon A
\end{equation}
at leading order in the small parameter $\varepsilon$, and neglecting the dynamics everywhere but on the $X$ nullcline.

\section{Periodic behaviour of the full two-gel system}
\begin{figure}
    \centering
    \input{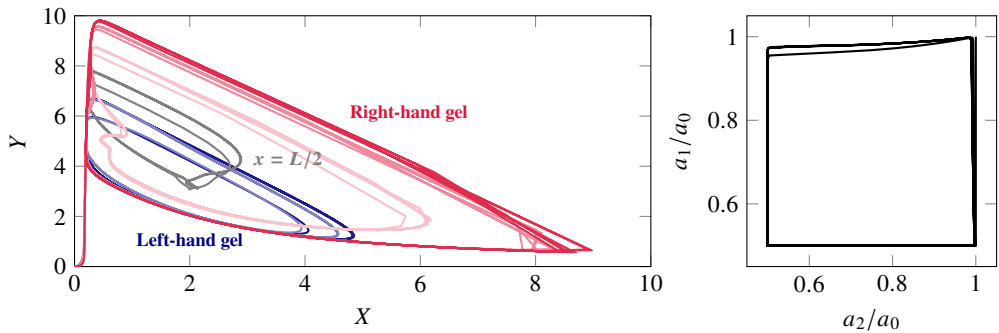}
    \caption{On the left, phase-plane plots of the full numerical solution to an oscillating BZ reaction coupled to a responsive gel with the same parameters as in figure 2 of the article, plotted at three equally-spaced points in both left- and right-hand gels and at the midpoint in the fluid, showing quasiperiodicity in all cases. On the right, plots of the quasiperiodic behaviour of the gel thickness as a result of these oscillations.}
    \label{fig:full_plot}
\end{figure}
If, instead of a simple spatially-constant oscillator, we consider two separate oscillators linked by diffusion and formulated as a coupled system of partial differential equations, the dynamics may be significantly different. The results presented in figure 2 of the article suggest a periodic nature in the concentration of chemical species $Y$ as the reaction progresses and the gels readjust accordingly. However, the dynamics illustrated here appear somewhat more complex than in a standard uncoupled Brusselator, with a non-constant (both in space and time) rate of increase in species $Y$ as the reaction rate inside the gels is modified, and potential loss of chemical species owing to diffusion. Figure \ref{fig:full_plot} plots trajectories in $X$-$Y$ space at various points within each gel and in the fluid between, showing that quasi-periodic behaviour is indeed seen as the oscillating reactions progress, but we cannot definitively prove the existence of limit cycles or the absence of chaos owing to the large dimensionality of the system and the lack of smoothness (when $Y$ crosses the threshold $Y^*$, the reaction rates change somewhat abruptly as a result of deswelling).

In spite of this, it is clear that there are choices of parameters in which stable (quasi-)periodic behaviour can persist. This is all that is required for our purposes -- when using signalling through frequency modulation, no distinction is made between the different forms of periodic solution, all that is required is an approximately constant time period and a slow decay rate in the amplitude of oscillation.

\section{Coupled relaxation oscillations in \texorpdfstring{$\mathbb{R}^4$}{R4}}
There has been some research into the dynamics of so-called `diffusively-coupled' Brusselator systems where one or both of the variables $X$ and $Y$ are linked with those of another oscillator through a term proportional to the difference in $X$ or $Y$ between the systems, but such examples tend to focus on identical chemical reactions and/or asymmetric coupling \citep{volkov-1995_bifurcations_brusselators,drubi-2007_coupling_chaos}. Regardless, it is apparent that rich dynamics including chaos can be exhibited in such cases: in \citet{mayora-cebollero-2025_almost_systems} two Brusselators are asymmetrically-coupled with identical reaction rates, and conditions for the systems to `almost synchronise' (for trajectories to remain close at most times) are deduced. In spite of synchronisation under strict conditions, certain parameter choices lead to strange attractor behaviour.

\begin{figure}
    \centering
    \input{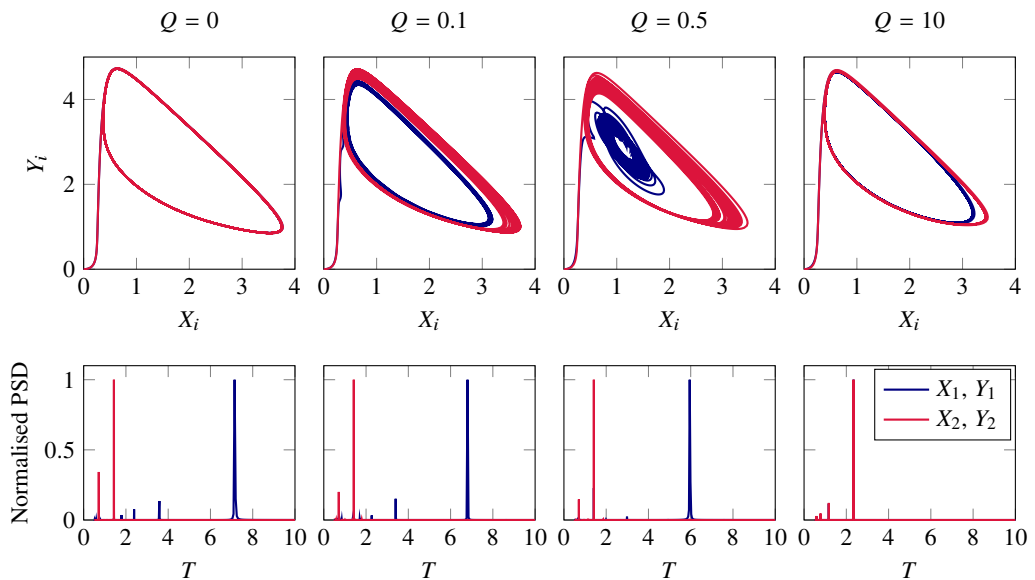}
    \caption{Phase-plane plots of the coupled Brusselator system with $A=1$, $B=3$ and $k=5$ as $Q$ is varied alongside a power spectral density plot (against period) showing the breakdown of periodic limit cycle dynamics when $Q=0$ into quasiperiodicity with potential chaos in the form of a strange attractor as $Q$ is increased, before settling into strongly coupled oscillators with a single shared period (and apparent limit cycle dynamics) as $Q \to \infty$.}
    \label{fig:chaos_plot}
\end{figure}

For the purposes of the present study, we consider symmetric coupling (of strength $Q$) but asymmetric reaction kinetics (one reaction occurs with a rate constant a factor $k$ greater), leading to the system
\begin{subequations}
    \begin{align}
        \dot{X}_1 &= A + X_1^2 Y_1 - (1+B)X_1 + Q(X_2-X_1),\\
        \dot{Y}_1 &= B X_1 - X_1^2 Y_1 + Q(Y_2-Y_1),\\
        \dot{X}_2 &= k\left[A + X_2^2 Y_2 - (1+B)X_2\right] + Q(X_1-X_2),\\
        \dot{Y}_2 &= k\left[B X_2 - X_2^2 Y_2\right] + Q(Y_1-Y_2),
    \end{align}
    \label{eqn:coupled_system}%
\end{subequations}
where dots represent time derivatives. Notice that this does not exactly replicate the coupled system in this article, where $k$ may in fact depend implicitly on $Y$ through deswelling or reswelling of the polymer scaffold, but this simpler model provides some good insight that can later be extended. In this case, we do not expect `almost synchronisation' in general, owing to the different reaction rates leading to wildly differing dynamics of each oscillator, but do qualitatively notice changing behaviour as $Q$ is increased, as shown in figure \ref{fig:chaos_plot}. This behaviour can be sorted into three regimes:
\begin{enumerate}
    \item When $Q=0$ the two oscillators behave independently as separate Brusselators each in $\mathbb{R}^2$, with corresponding limit cycles traversed at different rates.
    \item As $Q$ is increased, there is apparent quasiperiodicity for each oscillator, with the periods of each being in a fixed integer ratio.
    \item At large $Q$, the system functions as a single oscillator with $X_1=X_2$ and $Y_1=Y_2$.
\end{enumerate}
In some cases, we see more chaotic dynamics or notice that increasing $Q$ completely stabilises the fixed point and there is no periodic behaviour. Returning our attention to the fixed point $(X_i,\,Y_i) = (A,\,B/A)$, which remains a steady-state solution to the coupled system, the criteria for stability are modified from the $\mathbb{R}^2$ conditions, and are found by computing the eigenvalues of the Jacobian matrix
\begin{equation}
    J = \begin{pmatrix}B-(1+Q) & A^2 & Q & 0 \\ -B & -(A^2+Q) & 0 & Q \\ Q & 0 & k(B-1)-Q & k A^2 \\ 0 & Q & -k B & -kA^2 - Q\end{pmatrix}.
\end{equation}

Much has been written on the topic of bifurcations of coupled Brusselator systems, but we will henceforth consider cases where at least quasiperiodic behaviour (if not limit cycles) exists and the fixed points are unstable. We again assume $B \gg 1$ and so rewrite
\begin{subequations}
    \begin{align}
        \varepsilon \dot{X}_1 &= \varepsilon A + \varepsilon X_1^2 Y_1 - (1+\varepsilon)X_1 + \varepsilon Q(X_2-X_1),\\
        \varepsilon \dot{Y}_1 &= X_1 - \varepsilon X_1^2 Y_1 + \varepsilon Q(Y_2-Y_1),\\
        \varepsilon \dot{X}_2 &= k\left[\varepsilon A + \varepsilon X_2^2 Y_2 - (1+\varepsilon)X_2\right] + \varepsilon Q(X_1-X_2),\\
        \varepsilon \dot{Y}_2 &= k\left[X_2 - \varepsilon X_2^2 Y_2\right] + \varepsilon Q(Y_1-Y_2).
    \end{align}
    \label{eqn:coupled_system_scaled}%
\end{subequations}
In order to understand the dynamics of this system, we seek a perturbation series expansion of $X_i$ and $Y_i$ in powers of $\varepsilon$, taking
\begin{equation}
    X_i = X_{i0} + \varepsilon X_{i1} + \varepsilon^2 X_{i2} + \dots \qq{and} Y_i = Y_{i0} + \varepsilon Y_{i1} + \varepsilon^2 Y_{i2} + \dots,
\end{equation}
where it is easy to see that $X_{10} = X_{20} = 0$ and $X_{11} = X_{21} = A$. Then,
\begin{subequations}
    \begin{align}
        Y_{10} &= \frac{Y_{10}(0)+Y_{20}(0)}{2} + \frac{A(k+1)}{2}t - \frac{A(k-1)}{4Q}(1-e^{-2Qt}) + \frac{Y_{10}(0)-Y_{20}(0)}{2}e^{-2Qt}, \\
        Y_{20} &= \frac{Y_{10}(0)+Y_{20}(0)}{2} + \frac{A(k+1)}{2}t + \frac{A(k-1)}{4Q}(1-e^{-2Qt}) - \frac{Y_{10}(0)-Y_{20}(0)}{2}e^{-2Qt}.
    \end{align}%
\end{subequations}
Notice that, in the limit $Q\to 0$, these expressions reduce to exactly $Y_{10}(0)+At$ and $Y_{20}(0)+Akt$, reproducing the exact dynamics seen on the slow nullcline portion of the two individual limit cycles.

As $Q \to \infty$, we instead reach the limit where $Y_{10}=Y_{20}$ and the system behaves as a single Brusselator with effective rate constant $(1+k)/2$, as may be expected. Since the period of these oscillators scales with $\varepsilon^{-2}$, this is the case (for all but an initial transient) whenever $Q \gtrsim O(\varepsilon^2)$, since the exponentials will decay rapidly on the timescale of a single oscillation.

When the coupling is neither non-existent nor sufficiently strong as to lead to a single fully-synchronised oscillator, the faster of the two oscillators will potentially complete more than one cycle in the time it takes the slower to pass through one period. At the point where this fast cycle ends, the assumptions underpinning the perturbation expansion break down, and dynamics are fast, so we restrict our attention to the case $k > 1$ and consider the dynamics of the fast, second, oscillator within its period $0 < t < T_2$. Further assume that $Y_{10}(0)=Y_{20}(0) = 2 \varepsilon^{-1/2}$, as was seen in the single-oscillator system. Then, starting at $t=0$,
\begin{subequations}
    \begin{align}
        Y_{10} &= 2 \varepsilon^{-1/2} + \frac{A(k+1)}{2}t - \frac{A(k-1)}{4Q}(1-e^{-2Qt}), \\
        Y_{20} &= 2 \varepsilon^{-1/2} + \frac{A(k+1)}{2}t + \frac{A(k-1)}{4Q}(1-e^{-2Qt}).
    \end{align}%
\end{subequations}
These expressions allow us to find a more accurate estimate for the period of the faster oscillator by solving $Y_{20}(T_2)=(1+\varepsilon)^2/(4A\varepsilon^2)$ implicitly for $T_2$, with good agreement as shown in the first panel of figure \ref{fig:t1_t2_plot}. In the $n$\textsuperscript{th} period of the faster oscillator, the two $Y$ trajectories are described by
\begin{subequations}
    \begin{align}
        Y_{10}^{(n)} &= \frac{\hat{Y}^{(n-1)}}{2}(1+e^{-2Q\hat{t}}) + \frac{A(k+1)}{2}\hat{t} + \left(\varepsilon^{-1/2} - \frac{A(k-1)}{4Q}\right)(1-e^{-2Q\hat{t}}), \\
        Y_{20}^{(n)} &= \varepsilon^{-1/2}(1+e^{-2Q\hat{t}}) + \frac{A(k+1)}{2}\hat{t} + \left(\frac{\hat{Y}^{(n-1)}}{2} + \frac{A(k-1)}{4Q}\right)(1-e^{-2Q\hat{t}}),
    \end{align}%
    \label{eqn:y10y20}
\end{subequations}
\begin{figure}%
    \centering
    \input{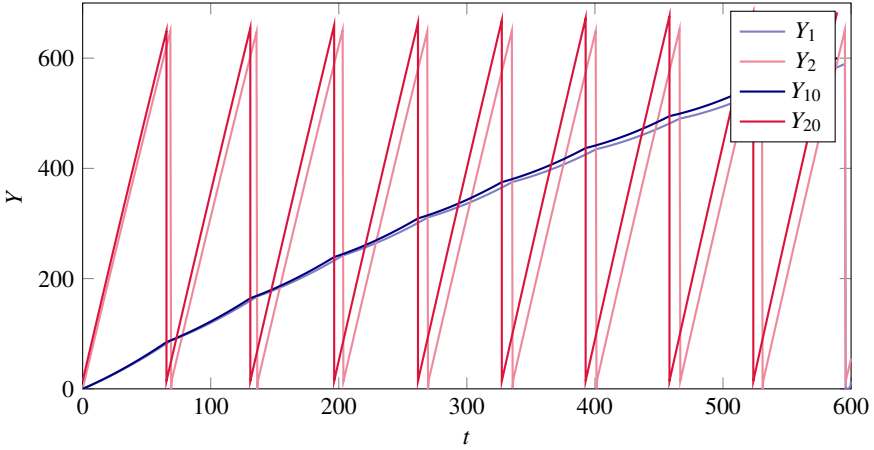}
    \caption{A comparison between the approximate forms of equations \eqref{eqn:y10y20} (dark curve) for $Y_1$ and $Y_2$ and the full numerical solution over one full period $T_1$, where $Q=2.5\varepsilon^2$ and all parameters are as in figure \ref{fig:t1_t2_plot}.}
    \label{fig:approx_compare_plot}%
\end{figure}%
where $\hat{t} = t - (n-1)T_2$ and $\hat{Y}^{(n)} = Y_{10}^{(n)}(T_2)$, and some sample solutions are plotted in figure \ref{fig:approx_compare_plot}, showing good agreement with the full solutions. This gives a recursion relation for the values of $Y_{10}$ at the end of the $n$\textsuperscript{th} oscillation of the fast oscillator, with
\begin{equation}
    \hat{Y}^{(n)} = 2\varepsilon^{-1/2} + \frac{A\left[2^n - \left(1+e^{-2 Q T_2}\right)^n\right]}{2^{n+1}}\left[e^{2 Q T_2}(1+k)\left(\coth{QT_2}-1\right)T_2 - \frac{A(k-1)}{Q} \right]
\end{equation}
At the end of the $n$\textsuperscript{th} period, the faster oscillator enters a region of fast dynamics, where $X_2$ rapidly increases, at the expense of $Y_2$, to a value $X_2 \approx (1+\varepsilon)^2/(4\varepsilon^2 A)$. The trajectory then intersects the $Y_2$ nullcline, on which
\begin{equation}
    k\left(X_2-\varepsilon^2 X_2^2 Y_2\right) = \varepsilon Q (Y_2-Y_1) \qq{so} \dot{X}_2 = k A - (k+Q) X_2 + k Q Y_1 + Q X_1.
\end{equation}
At early times after reaching this nullcline, the dynamics can be approximated by $\dot{X}_2 \approx -k X_2$, since $Q$ is small and all other terms are insignificant, so $X_2 \approx (1+\varepsilon)^2(4\varepsilon^2 A)^{-1} e^{-k \hat{t}}$. When $X_2$ increases quickly, $X_1$ also does so, with rapid dynamics described approximately by
\begin{align}
    \dot{X}_1 \approx Q X_2 \qq{so} X_1 &\approx \frac{1+\varepsilon - \sqrt{1-4A\varepsilon^2 \hat{Y}^{(n)}}}{2\varepsilon \hat{Y}^{(n)}} + \frac{(1+\varepsilon)^2 Q}{4A\varepsilon^2 k}(1-e^{-k\hat{t}}) \notag \\ &\approx \frac{1}{2\hat{Y}^{(n)}} + \frac{(1+\varepsilon)^2Q}{4A\varepsilon^2 k}(1-e^{-k\hat{t}}).
\end{align}
Therefore, provided that
\begin{equation}
    \hat{Y}^{(n)} > \frac{1/\varepsilon + 1 + Q}{X_1} - \frac{A + Q(1+\varepsilon)^2(4\varepsilon^2 A)^{-1}e^{-k\hat{t}}}{X_1^2}
\end{equation}
for all times $t$, $X_1$ also increases and jumps rapidly to join the $\dot{Y}_1 = 0$ nullcline, completing a period. We can restate this in the form of the criterion
\begin{equation}
    \hat{Y}^{(n)} > \frac{A}{Q+4A^2\varepsilon^2}\left[1 + 2(1+k)\varepsilon + (k^2-4k+1)\varepsilon^2\right] + \dots
    \label{eqn:inequality}
\end{equation}
Then, the period of the slower oscillator, $T_1$, is given by $T_1=nT_2$ for the lowest $n$ that satisfies inequality \eqref{eqn:inequality}, unless this would lead to a value of $\hat{Y}^{(n)}$ greater than that expected for an uncoupled oscillator, in which case $n=k$. Figure \ref{fig:t1_t2_plot} shows how well these predictions correspond to the full numerical solutions, illustrating how the periods reach a regime where they are integer multiples of one another before settling towards $T_1=T_2$.
\begin{figure}
    \centering
    \input{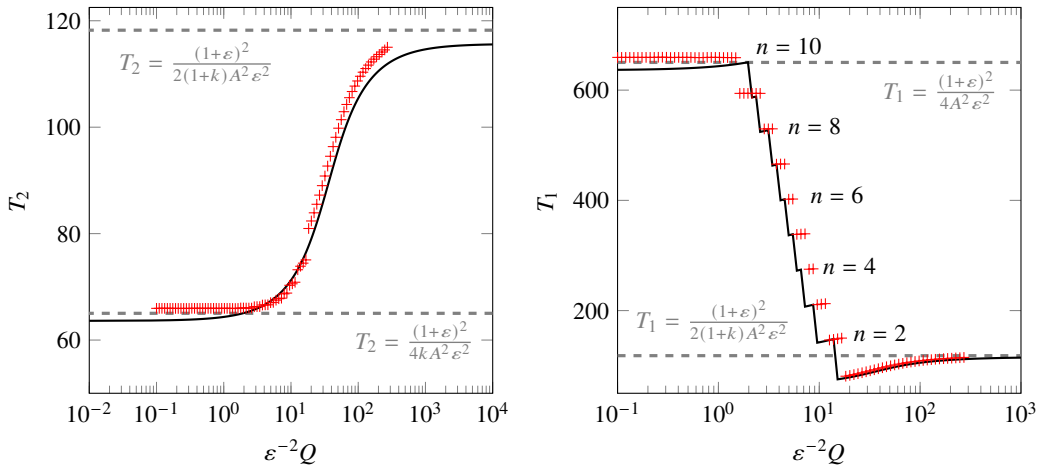}
    \caption{Plots of the period of the faster oscillator ($T_2$) and the slower oscillator ($T_1$) in a system with $A=1$ and $\varepsilon = 1/50$ with $k=10$ as $Q$ is varied, with exactly-computed numerical solutions plotted in red. The approximation to $T_2$ is computed using equation \eqref{eqn:y10y20}, whilst the expression for $T_1$ is determined in terms of the predicted value of $T_2$ using expression \eqref{eqn:inequality}. Beyond $Q \approx 0.12$, the numerical solutions become stable and coupling damps away any periodic behaviour.}
    \label{fig:t1_t2_plot}
\end{figure}

\subsection{Rate-dependence on \texorpdfstring{$\Phi$}{polymer fraction}}
In reality, the coupled system we have considered so far in the present section is not an exact model for the system of equation (10) where the reaction rate constant depends on the degree to which the gels are swollen. Instead of equations \eqref{eqn:coupled_system}, the governing equations for this system are
\begin{subequations}
    \begin{align}
        \dot{X}_1 &= \Phi_1\left[A + X_1^2 Y_1 - (1+B)X_1 + Q(X_2-X_1)\right],\\
        \dot{Y}_1 &= \Phi_1\left[B X_1 - X_1^2 Y_1 + Q(Y_2-Y_1)\right],\\
        \dot{X}_2 &= \Phi_2\left[A + X_2^2 Y_2 - (1+B)X_2 + Q(1-E)(X_1-X_2)\right],\\
        \dot{Y}_2 &= \Phi_2\left[B X_2 - X_2^2 Y_2 + Q(1-E)(Y_1-Y_2)\right],
    \end{align}
    \label{eqn:coupled_system_phi}%
\end{subequations}
Working in the $Q \to \infty$ limit where $X_1=X_2$ and $Y_1 = Y_2$, this system becomes equivalent to a single oscillator in $\mathbb{R}^2$ with a rate constant given by $(\Phi_1+\Phi_2)/2$.

Below $Y=Y^*$, this represents a Brusselator with rate constant $(1+\Phi_2^-)/2$, whilst above it the system becomes a Brusselator with rate constant $(\Phi_\infty+\Phi_2^+)/2$. An equivalent calculation to equation \eqref{eqn:time_period_calc} thus allows us to deduce
\begin{align}
    T &\approx \int^{Y^*}_{2/\varepsilon^{1/2}}{\frac{2}{(1+\Phi_2^-)A}\,\mathrm{d}Y} + \int_{Y^*}^{(1+\varepsilon)^2/4\varepsilon^2 A}{\frac{2}{(\Phi_\infty+\Phi_2^+)A}\,\mathrm{d}Y} \notag \\
    &\approx  \frac{(1+\varepsilon)^2}{2 A^2 \varepsilon^2 (\Phi_\infty+\Phi_2^+)} + \frac{2 Y^*}{A}\left(\frac{1}{1+\Phi_2^-} - \frac{1}{\Phi_\infty+\Phi_2^+}\right),
\end{align}
whenever $\varepsilon = 1/B \ll 1$ and $\varepsilon^{-1/2} \ll Y^*$. If, further, $Y^* \ll \varepsilon^{-2}$,
\begin{equation}
    T \approx  \frac{(1+B)^2}{4A^2}\frac{2}{\Phi_\infty+\Phi_2^+}.
\end{equation}

\bibliographystyle{jfm}